\documentclass[a4paper,12pt]{article}
\usepackage{graphicx,amssymb,bm,latexsym,color,epsf,cite}
\makeatletter
\@addtoreset{equation}{section}

\makeatother
\newcommand{\be}{\begin{equation}}
\newcommand{\ee}{\end{equation}}
\newcommand{\bea}{\begin{eqnarray}}
\newcommand{\eea}{\end{eqnarray}}
\newcommand{\vs}[1]{\vspace{#1 mm}}

\renewcommand{\a}{\alpha}
\renewcommand{\b}{\beta}
\renewcommand{\c}{\gamma}

\renewcommand{\d}{\delta}

\newcommand{\la}{\lambda}
\newcommand{\pa}{\partial}

\newcommand{\dsl}{\pa \kern-0.5em /}
\newcommand{\Dsl}{D \kern-0.65em /}
\newcommand{\nn}{\nonumber\\}
\newcommand{\p}[1]{(\ref{#1})}

\newcommand{\cL}{{\cal L}}

\newcommand{\bnabla}{\bar\nabla}
\newcommand{\br}{\bar R}
\newcommand{\bg}{\bar g}

\newcommand{\Det}{{\rm Det}}

\begin{document}
\renewcommand{\thefootnote}{\fnsymbol{footnote}}

\begin{flushright}
KU-TP 076 \\
\today
\end{flushright}
\vs{5}

\begin{center}
{\Large\bf General Procedure of Gauge Fixings and Ghosts}
\vs{10}

{\large
Nobuyoshi Ohta\footnote{e-mail address: ohtan@phys.kindai.ac.jp}
} \\
\vs{10}

{\em Department of Physics, Kindai University,
Higashi-Osaka, Osaka 577-8502, Japan}

\vs{10}
{\bf Abstract}
\end{center}

We revisit the general procedure of gauge fixings and ghosts based on BRST invariance principle.
It is shown that when this is applied to the higher-derivative gauge fixings,
it gives the correct structure of gauge fixings and ghosts including ``third ghost'',
previously derived at one-loop level. This procedure is solely based on the symmetry principle
and is valid at full order.

\renewcommand{\thefootnote}{\arabic{footnote}}
\setcounter{footnote}{0} 

\section{Introduction}

The necessity of ghosts in the quantization of gravity was first pointed out by Feynman~\cite{Feynman},
and subsequently formulated by DeWitt~\cite{DeWitt}.
Faddeev and Popov gave a lucid derivation of the ghosts in the path integral approach
to the quantization of gauge theories~\cite{FP}, named FP ghosts.
The gauge-fixing procedure breaks the (classical) gauge invariance of the system, and
the ghost is necessary to have unitarity of the theory.

Later it was recognized that the resulting system with gauge fixing and ghosts has an invariance
under Fermionic transformation, called BRST transformation~\cite{BRS}.
This is a recovery of the gauge invariance at the quantum level, and it turned out to be quite useful in proving
renormalizability~\cite{BRS} and unitarity~\cite{KO} in nonabelian gauge theory as well as gravity~\cite{stelle}.
Further the BRST invariance was used as a proper way of gauge fixing in skew-symmetric tensor gauge
theories in which ghosts for ghosts necessarily appear in sequence~\cite{Kimura,HKO}.
In these examples, the original FP path integral method does not work, or is very difficult to apply
if not impossible. The procedure was neatly summarized in~\cite{KU}.

There are other circumstances in which the method is quite useful.
An example is the quantization of gauge theories and gravity with higher-derivative gauge fixings.
In the gravity case, it was pointed out long ago that we need ``third'' ghost in addition to the known FP ghosts
when we have higher derivatives~\cite{BC}.
The functional-integral representation of the one-loop part of the vacuum-to-vacuum amplitude is given by
\bea
e^{iW_{(1)}} = N (\det Y_{\a\b})^{1/2}\int \exp\left[\frac{i}{2} \phi^i F_{ij}\phi^j \right] [\det F^\a{}_\b]d\phi^i,
\label{eff1}
\eea
in DeWitt's condensed notation, where
\bea
F_{ij} \equiv  S_{,ij} +\phi^i P^\a{}_i Y_{\a\b} P^\b{}_j,
\eea
is the second variation of the total action with the linear gauge fixing term
\bea
P^\a{}_i \phi^i = \zeta^\a,
\eea
with some differential operator $P^\a{}_i$ and a constant $\zeta^\a$, and $Y_{\a\b}$ is a matrix operator
symmetric and nonsingular.
The operator $F^\a{}_\b$ is constructed as
\bea
F^\a{}_\b =P^\a{}_i Q^i{}_\b.
\eea
Here the operator $Q^i{}_\a$ is defined by the gauge transformation
\bea
\d\phi^i =Q^i{}_\a \d\xi^\a,
\eea
under which the theory is invariant, with the infinitesimal gauge-group parameter $\d\xi^\a$.
We can integrate \p{eff1} to obtain
\bea
e^{iW_{(1)}} = N \frac{(\det Y_{\a\b})^{1/2}\, [\det F^\a{}_\b ]}{(\det F_{ij})^{1/2}}.
\label{eff2}
\eea
The term in the denominator gives the contribution of the physical particles of the system.
The factor $[\det F^\a{}_\b ]$ gives the well-known FP ghost, but the first factor $(\det Y_{\a\b})^{1/2}$
gives less familiar ``third'' ghost~\cite{BC}.
If $Y_{\a\b}$ is $\eta_{\a\b}$, this does not contribute to the effective action and can be dropped.

Though the origin of the ghosts may be traced back to the path integral formalism,
we have to go through the complicated procedure to confirm what is the right action.
We should also note that the above formulation is valid only at the one-loop level,
and it is not clear how to modify the ghosts beyond one loop.
Here we point out that there is a much simpler universal procedure solely from the viewpoint of symmetry principle,
the BRST invariance. We can easily obtain the correct result without detailed consideration.
Moreover, this formulation, being solely based on the symmetry principle, gives ghost terms valid at full level.

The method was also applied to quantization of gravitino in supergravity~\cite{HK}.
We also briefly review how this method gives a concise and clear derivation of
the necessary ghosts and their kinetic terms.
However we point out that if we consider the theory in curved spacetime,
there is an important contribution dropped in \cite{HK}.

Before getting into the concrete examples, let us first summarize the general procedure.

\section{General procedure of gauge fixing}

Here we first summarize the procedure how to write down the BRST invariant gauge-fixing and
ghost terms.

The BRST transformation is identified in the following way.
\begin{enumerate}
\item 
Given the (classical) gauge transformation, replace the gauge parameter by an anti-commuting ghost $c$.

In the case of gauge theory, we have the infinitesimal transformation
\bea
\d A^i(x) = i \Lambda^a(x) T_a^i[A],
\label{trans1}
\eea
where $A^i(x)$ are the gauge or matter fields, $\Lambda^a(x)$ are the gauge parameters,
and $a$ is an index of Lie group. We then have the BRST transformation
\bea
\d_B A^i = i\d\la\, c^a(x) T_a^i[A],
\label{trans2}
\eea
where $c^a(x)$ is a FP ghost, and $\d\la$ is a global Grassmann parameter.

\item
The BRST transformation of the ghost is determined by the requirement that the BRST transformation~\p{trans1}
be nilpotent.

For gauge theory, this determines
\bea
\d_B c^a(x) = \frac{1}{2} f_{bc}{}^a\, c^b(x) c^c(x),
\label{trans3}
\eea
where $f_{bc}{}^a$ is the structure constant of the group.

\item
Finally the BRST transformations of the antighost and associated auxiliary field $B_a(x)$ are defined as
\bea
\d_B \bar c_a(x) &=& i \d\la\, B_a(x), \nn
\d_B B(x) &=& 0.
\eea

\end{enumerate}
We see from \p{trans1} that this transformation is reminiscent of the original (classical) gauge transformation;
the only nontrivial transformation is \p{trans3} which is necessary for the invariance of the whole system.
We also note that the BRST transformation is defined prior to any particular gauge-fixing conditions.
This is made possible by the introduction of the {\it auxiliary} field $B$.
We will see that this field $B$ will play more important roles in the higher-derivative gauge fixings.

The BRST invariant gauge-fixing and ghost terms are given by~\cite{Kimura,HKO,KU}
\bea
{\cal L}_{\rm GF+FP} &=& -i \d_B\left[\bar c_a \left(F^a-\frac{\a}{2} B^a\right) \right]/\d\la \nn
&=& B_a F^a - \frac{\a}{2} B_a B^a + i \bar c_a (\d_B F^a) /\d\la,
\eea
where $\a$ is the gauge fixing parameter and $F^a$ is the gauge fixing function.
Eliminating the $B$ field, we obtain a familiar form
\bea
{\cal L}_{\rm GF+FP}
= \frac{1}{2\a} (F_a)^2 + i \bar c_a (\d_B F^a)/\d\la.
\eea
It is clear from the nilpotency of the BRST transformation that this gauge-fixing and ghost terms are
automatically BRST invariant.
This ensures that the additional gauge modes decouple from the physical subspace,
and we obtain gauge invariant result.

We emphasize that this procedure is quite general; for example, the gauge-fixing function $F^a$
may contain higher-derivative terms. It may even include ghosts and we can find the correct terms.
This is difficult in the path integral approach if not impossible.
In the next section, we apply this to higher-derivative gauge fixing and show that it naturally leads to
the ``third'' ghost.

\section{Higher-derivative gauge fixing and ghosts}
\label{ex}

\subsection{Gauge fixing for higher-derivative gravity}

When the gravity theory contains higher derivatives, it is natural to have gauge fixing with higher derivatives.
We split the metric $g_{\mu\nu} = \bg_{\mu\nu} + h_{\mu\nu}$, 
where $\bg_{\mu\nu}$ is an arbitrary background.
The BRST transformation is given by
\bea
&& \d_B h_{\mu\nu} = -\d\la (\nabla_\mu c_\nu + \nabla_\nu c_\mu), \nn
&& \d_B c^\mu = \d\la c^\rho \pa_\rho c^\mu,~~
\d_B \bar c_\mu = i\d\la B_\mu, ~~
\d_B B_\mu=0,
\label{ggt}
\eea
where $c^\mu, \bar c_\mu$ and $B_\mu$ are FP ghost, anti-ghost and auxiliary fields
with Grassmann odd parameter $\d\la$.
The covariant derivative in \p{ggt} is full one, and it would lead to general ghost terms.
However, when we compute the effective action for vanishing expectation value of $h_{\mu\nu}$, or when we consider
one-loop effects, we can restrict this to the linear terms:
\bea
\d_B h_{\mu\nu}
= -\d \la ( \bnabla_\mu c_\nu + \bnabla_\nu c_\mu),
\label{ggt2}
\eea
where the barred covariant derivative $\bnabla$ is constructed with the background metric.
We can write the gauge-fixing and FP terms as
\bea
{\cal L}_{GF+FP}/\sqrt{\bg}
&=& -i\d_B \left[\bar c_\mu Y^{\mu\nu} \left(F_\nu-\frac{\a}{2}B_\nu\right)\right]/\d\la \nn
&=& B_\mu Y^{\mu\nu} F_\nu - \frac{\a}{2}B_\mu Y^{\mu\nu} B_\nu
+i \bar c_\mu Y^{\mu\nu}\Delta^{(gh)}_{\nu\rho}c^\rho,
\label{gfgh}
\eea
where
\bea
F_\mu &\equiv& \bnabla^\la h_{\la\mu} + \b \bnabla_\mu h\ , \nn
\Delta^{(gh)}_{\mu\nu} &\equiv & g_{\mu\nu} \bnabla^2 +(2\b+1)\bnabla_\mu \bnabla_\nu +\br_{\mu\nu}\ , \nn
Y_{\mu\nu} &\equiv& \bg_{\mu\nu} \bnabla^2+ \c \bnabla_\mu \bnabla_\nu - \xi \bnabla_\nu \bnabla_\mu\ ,
\eea
where $\a$, $\b$, $\c$ and $\xi$ are gauge parameters.
We can rewrite \p{gfgh} as
\bea
{\cal L}_{GF+FP}/\sqrt{\bg}
= \frac{1}{2\a} F_\mu Y^{\mu\nu} F_\nu - \frac{\a}{2} \tilde B_\mu Y^{\mu\nu} \tilde B_\nu
+i \bar c_\mu Y^{\mu\nu}\Delta^{(gh)}_{\nu\rho}c^\rho,
\label{gfgh2}
\eea
with
\bea
\tilde B_\mu=B_\mu-\frac{1}{\a} F_\mu.
\eea
We see that we get the same correct determinant factor $(\det Y^{\mu\nu})^{1/2}$
as \p{eff2} after we perform the path integral over $B_\mu$ and FP ghosts.
It is important to note that the additional mode appears here from the original ``auxiliary'' field $B_\mu$;
the original ``auxiliary'' field is no longer auxiliary in the higher-derivative gauge fixing.
This gives a new view of the origin of the ``third'' ghost solely from the symmetry principle.

Eq.~\p{gfgh} is the form of gauge-fixing used, say in \cite{OP2013}.
Higher-derivative gauge fixing was also used in \cite{stelle} to discuss renormalization of
higher-derivative gravity, but such additional ghost was not considered.
This is not a problem because there the discussions are restricted to the flat backgrounds,
and the additional ghost is free and can be neglected.
However we emphasize that the additional ghosts are very important in the nontrivial backgrounds.

We have restricted the above discussions to one-loop case, but this is just for simplicity and
for comparison with earlier results. It is straightforward to get the full structure just
by using the full transformation~\p{ggt} instead of \p{ggt2} to derive expressions valid to all orders.
It should be possible to derive the same functional determinants in the usual path integral approach
for the full case, as was done in Ref.~\cite{BC} at one loop. The above coincidence of the results
at one loop gives justification of our results to some extent. Here we take the viewpoint that
this gives an alternative formulation solely based on the principle of gauge (BRST) invariance
at the quantum level, and clearly explains the origin of the ``third'' ghost.

Though the higher-derivative terms in \p{gfgh} or \p{gfgh2} give bona fide formulation in the path integral
quantization, they sometimes cause subtlety in the canonical operator formalism.
It is more convenient to transform them into a first-order ghost theory as follows.
With a new antighost defined by
\bea
\bar c^\mu{}' = Y^{\mu\nu}\bar c^\nu,
\eea
the BRST transformation for the antighost $\bar c^\mu{}'$ is modified as
\bea
\d_B \bar c^\mu{}' = i\d\la  Y^{\mu \nu}B_\nu.
\eea
Then the gauge-fixing and FP ghost terms~\p{gfgh} reduce to
\bea
{\cal L}_{GF+FP}'/\sqrt{\bg}
&=& -i\d_B \left[\bar c^\nu{}' \left(F_\nu-\frac{\a}{2}B_\nu\right)\right]/\d\la \nn
&=& F_\mu Y^{\mu\nu} B_\nu -\frac{\a}{2}B_\mu Y^{\mu\nu} B_\nu
+i \bar c^\mu{}' \Delta^{(gh)}_{\nu\rho}c^\rho.
\label{gfgh3}
\eea
Since when we integrate out $B$, $c$ and $\bar{c}$, the path integrals with \p{gfgh} and \p{gfgh2}
differ by a determinant, we have to include in addition to the gauge fixing and ghosts an additional term
\bea
{\cal L}_{\rm aux}/\sqrt{\bg}  = i \bar{d}^{\mu} Y_{\mu\nu} d^{\nu},
\eea
where $ \bar{d}^{\mu} $ and $d^{\mu}$ are anti-commuting independent fields and are BRST invariant.

We can redefine the ``auxiliary'' field $B_\mu$ in Eq.~\p{gfgh3}, and the whole Lagrangian becomes
\bea
{\cal L}_{GF+FP}''/\sqrt{\bg}
= \frac{1}{2\a} F_\mu Y^{\mu\nu} F_\nu -\frac{\a}{2} \tilde B_\mu Y^{\mu\nu} \tilde B_\nu
+i \bar c_\mu{}' \Delta^{(gh)}_{\nu\rho}c^\rho + i \bar{d}^{\mu} Y_{\mu\nu} d^{\nu}.
\eea
Note again that the kinetic term for ``auxiliary'' field $\tilde B_\mu$ contains the factor $Y^{\mu\nu}$
and $\tilde B_\mu$ is dynamical.
This form was used in \cite{FOP}.

\subsection{Gravitino}

Another circumstance in which the formalism is useful is the gravitino gauge fixing for supersymmetry
with an additional ghost degree of freedom~\cite{HK}.
Conventionally the gauge-fixing and the FP ghost were taken as
\bea
\cL_{GF}+\cL_{FP} = \frac{i}{2\a}(\bar\psi\cdot\c)\dsl(\c\cdot\psi)+\bar c_* \Dsl c,
\eea
where $\psi_\mu$ is a Majorana gravitino, $\c\cdot\psi\equiv \c_\mu \psi^\mu$, and $c$ and $c_*$ are
the commuting spinor ghost and antighost, respectively.
However it was first pointed out in \cite{Nielsen:1978,Kallosh:1978} and elaborated in a beautiful manner
in \cite{HK} using the BRST transformation that these terms should be
\bea
\cL_{GF}+\cL_{FP} = \frac{i}{2\a}(\bar\psi\cdot\c)\dsl(\c\cdot\psi)-i\bar c_* \dsl \Dsl c,
\label{ggffp1}
\eea
for the unitarity of the theory.
Note the extra factor $\dsl$ in the last term, which introduces an additional mode.
This could be derived as the BRST invariant form
\bea
\cL_{GF}+\cL_{FP} = - \d_B \left[ \bar c_* \dsl\left(\c\cdot\psi +i \frac{\a}{2} B\right)\right]/\d\la,
\label{ggffp2}
\eea
where it is understood that background metric (which is inert under the BRST transformation)
is used in the gauge fixing term and the BRST transformation restricted to supersymmetry is given by
\bea
\d_B \psi^\mu &=& i \d\la D^\mu c,
\nn
\d_B c &=& \frac{1}{4}\d\la\, \psi_\mu \bar c\c^\mu c,
\nn
\d_B c_* &=& \d\la B,
\nn
\d_B B &=& 0,
\eea
with a Grassmann parameter $\d\la$. Then \p{ggffp2} in fact yields
\bea
\cL_{GF}+\cL_{FP} = -\bar B \dsl(\c\cdot\psi)-\frac{i}{2}\a \bar B\dsl B - i\bar c_* \dsl \Dsl c,
\eea
which gives \p{ggffp1} after integrating out the field $B$, modulo the determinant $\Det(\dsl)$ coming from
the path integral of $B$ field. This was not incorporated in \cite{HK}. As in the case of \cite{stelle},
this is not a problem on the flat spacetime. However if we consider the theory in the curved spacetime,
the simple derivative should be replaced with the covariant derivative, and this factor gives important contribution.
Once again, note that the ``auxiliary'' field $B$ is no longer auxiliary in the presence of the higher-derivative
gauge fixing.

\section*{Acknowledgment}

We would like to thank Kevin Falls and Taichiro Kugo for valuable discussions.
This work was supported in part by the Grant-in-Aid for Scientific Research Fund of the JSPS (C) No. 16K05331
and 20K03980.


\end{document}